\documentstyle[12pt]{article}

\def\L{\Lambda}\def\t{\tau}
\def\bR{{\bf R}}\def\bT{{\bf T}}
\def\sR{{\sf R}}\def\sT{{\sf T}}
\def\S{\Sigma}

\begin{document}
\advance\textheight 12pt

\begin{center}
\large{{\bf Spherically Symmetric Scalar Field Collapse:
An Example of the Spacetime Problem of Time}}
\end{center}

\vspace{20pt}
\begin{center}
by\\
Joseph D. Romano\\
Department of Physics\\
University of Utah\\
Salt Lake City, UT 84112
\end{center}

\section*{Abstract}

A canonical formalism for spherical symmetry, originally developed by
Kucha\v{r} to describe vacuum Schwarzschild black holes, is extended to
include a spherically symmetric, massless, scalar field source.
By introducing the ADM mass as a canonical coordinate on phase space, one
finds that the super-Hamiltonian and supermomentum constraints for the
coupled system simplify considerably.
Yet, despite this simplification, it is difficult to find a functional
time formalism for the theory.
First, the configuration variable that played the role of time for the
vacuum theory is no longer a spacetime scalar once spherically symmetric
matter is coupled to gravity.
Second, although it is possible to perform a canonical transformation to
a new set of variables in terms of which  the super-Hamiltonian and
supermomentum constraints can be solved, the new time variable also fails
to be a spacetime scalar.
As such, our solutions suffer from the so-called {\it spacetime problem of
time}.
A candidate for a time variable that {\it is} a spacetime scalar is
presented.
Problems with turning this variable into a canonical coordinate on phase
space are discussed.

\bigskip
\noindent PACS number(s): 0420, 0460, 9760L

\section*{1. Introduction}

Canonical quantization is well-suited for the study of collapsing matter
systems.
First, by quantizing both geometry and matter, canonical quantization goes
beyond the semi-classical approximation used, for example, in the standard
treatment of Hawking radiation\cite{Hawking}.
Second, by working on arbitrary Cauchy hypersurfaces, one can study what
happens to the canonical data {\it inside a black hole} as one approaches
the curvature singularity.
Third, by performing a midisuperspace reduction to spherically symmetric
spacetimes, one obtains simpler models that, hopefully, one can then solve.
Canonical quantization is thus a promising method for investigating the
formation and evaporation of black holes, and for studying the nature of
horizons and singularities in quantum theory.

As a first step toward obtaining a better understanding of spherically
symmetric gravitational collapse, Kucha\v{r}\cite{Kuchar1} has given a
detailed and elegant analysis of the canonical quantization of vacuum
Schwarzschild black holes.
He was able to cast the classical and quantum dynamics of primordial
black holes into a simple and geometrically transparent form by turning
the curvature radius $R$ and Killing time $T$ of the Schwarzschild solution
into canonical coordinates on the geometrodynamical phase space:
\begin{equation}
m,p\,;\,\,\,\sT(r),P_\sT(r)\,;\,\,\,\sR(r),P_\sR(r)\,.
\label{eq:chart}
\end{equation}
$T$ and $R$ thus become embedding variables $\sT(r)$ and $\sR(r)$ that
specify how the Cauchy hypersurfaces are drawn in the spacetime.
The canonical variables $m$ and $p$ also have a simple physical meaning:
$m$ is the Schwarzschild mass of the spacetime, and $p$ is the difference
between proper times at the right and left infinities.

In terms of these canonical variables, the super-Hamiltonian and
supermomentum constraints are equivalent to
\begin{equation}
P_\sT(r)=0\,,\quad P_\sR(r)=0\,.
\label{eq:vacconstraints}
\end{equation}
The Hamiltonian, which is a linear combination of these constraints,
weakly vanishes, implying that the canonical variables $m$ and $p$ are
constants of motion.
The Dirac quantization of this theory is also particularly simple.
Wave functions $\Psi=\Psi(m,t;\sT,\sR]$ satisfying the quantum version of
the constraints (\ref{eq:vacconstraints}) are independent of $\sT(r)$ and
$\sR(r)$.
Since the Hamiltonian of the theory vanishes, wave functions are also
independent of the label time $t$.
The final result: $\Psi=\Psi(m)$.

The next step is to extend the above analysis to include a spherically
symmetric, massless, scalar field source.

To some extent, the geometrodynamics of a spherically symmetric, massless,
scalar field coupled to gravity has already been worked out.
Berger, Chitre, Moncrief, and Nutku (BCMN)\cite{BCMN} addressed this problem
in the early 1970's.
Subsequently, Unruh\cite{Unruh} and H\'aj\'{\i}\v{c}ek\cite{Hajicek}
carefully analyzed the BCMN model, especially in regard to black hole
evaporation and the properties of apparent horizons in the canonical
formalism.

But in all of these treatments, the action for the coupled system is
{\it reduced to a privileged foliation} specified by the vanishing of the
``radial'' momentum.
The Cauchy hypersurfaces are no longer arbitrary; they are selected by
the above slicing condition.
For the vacuum theory, this slicing condition amounts to working on the
surfaces of constant Killing time $T$.
These hypersurfaces thus cover only the static regions of the Kruskal diagram
and fail to penetrate the horizon.
H\'aj\'{\i}\v{c}ek\cite{Hajicek} also chooses to foliate spherically
symmetric spacetimes in such a way that the region interior to an apparent
horizon is removed.

This is {\it not} what we want to do.

Rather, we want to be able to choose Cauchy hypersurfaces so that we
can study what happens to the canonical data {\it inside a black hole} as we
approach the curvature singularity.
As such, we need our foliation to cover the whole spacetime; the
hypersurfaces must be able to penetrate an apparent horizon.
In addition, we want to know how the hypersurfaces are located in the
spacetime.
This means that we need to have embedding variables as canonical coordinates
on phase space.
Given that the super-Hamiltonian and supermomentum constraints can then
be solved for the momenta canonically conjugate to these variables,
the Dirac quantization of the theory would be described by wave functions
satisfying first-order functional Schr\"odinger equations.
In this way, we would avoid the difficulties associated with solving the
second-order Wheeler-DeWitt equation.

In other words, we desire a {\it functional time formalism} for our
collapsing matter system.

As mentioned earlier, Kucha\v{r}\cite{Kuchar1} succeeded to find a
functional time formalism for vacuum spherically symmetric spacetimes.
The purpose of this paper is to present two attempts to find a functional
time formalism for the coupled system, and to show how these attempts fail.
Basically, the solutions suffer from the so-called {\it spacetime problem
of time}\cite{Kuchar2}.
That is, the time variables that we introduce as canonical coordinates on
phase space are not spacetime scalars, and hence fail to qualify as true
embedding variables.

The plan of the rest of the paper is as follows:
In section 2, we briefly describe the canonical formalism for a spherically
symmetric, massless, scalar field coupled to gravity.
In section 3, we introduce the ADM mass as a canonical coordinate on phase
space, thereby simplifying the constraints just as Kucha\v{r} did for the
vacuum theory.
In section 4, we define what we mean by the spacetime problem of time
and show that the time variable $\sT(r)$, originally introduced for the
vacuum theory, is not a spacetime scalar once spherically symmetric matter
is coupled to gravity.
We also point out that, although one can introduce a new time variable
$\bT(r)$ in terms of which we can solve the super-Hamiltonian and
supermomentum constraints, $\bT(r)$ also fails to be a spacetime scalar.
Finally, in section 5, we conclude by presenting a natural candidate for a
time variable that {\it is} a spacetime scalar---the curvature time $T$ of
the general, spherically symmmetric, spacetime line element---and discuss
the problems of turning this privileged spacetime coordinate into a
canonical coordinate on phase space.

\section*{2. Canonical formalism}

Let $(\S,g)$ be a 3-dimensional, spherically symmetric, Riemannian space
with coordinates $x^{a}=(r, \theta, \phi)$ adapted to the symmetry.
The line element $d\sigma$ on $\S$ can be written as
\begin{equation}
d\sigma^2 = \L^2(r)\,dr^2 + R^2(r)\,d\Omega^2
\label{eq:dsigma2}
\end{equation}
where $d\Omega^2:=d\theta^2+\sin^2\theta\,d\phi^2$ is the line element on
the unit 2-sphere.
Note that $d\sigma$ is completely characterized by two functions $\L(r)$ and
$R(r)$ of the radial label $r\in[0,\infty)$.
The point $r=0$ is the center of spherical symmetry.

Modulo boundary terms, the vacuum dynamics of the gravitational field
follows from the the ADM action
\begin{equation}
S^G:=\int dt\int_0^\infty dr\,\left(P_\L\dot\L + P_R\dot R - NH^G-N^rH^G_r
\right)
\label{eq:SG}
\end{equation}
where
\begin{eqnarray}
H^G&:=&-R^{-1} P_R P_\L + {1\over 2}R^{-2} \L P_\L^2\nonumber\\
& & \mbox{} +\L^{-1} R R''-\L^{-2} RR' \L' + {1\over 2}\L^{-1} R'{}^2
- {1\over 2}\L
\label{eq:HG}\\
H^G_r&:=& P_R\,R' - \L P_\L'
\label{eq:HGr}
\end{eqnarray}
are the gravitational super-Hamiltonian and supermomentum.
The dynamics of a spherically symmetric, massless, scalar field propagating
on this spacetime follows from the action
\begin{equation}
S^\phi:=\int dt\int_0^\infty dr\,\left(\pi\dot\phi - NH^\phi-N^r H_r^\phi
\right)
\label{eq:Sphi}
\end{equation}
where
\begin{eqnarray}
H^\phi&:=&{1\over 2}\, \L^{-1}\left(R^{-2}\pi^2+ R^2\phi'{}^2\right)
\label{eq:Hphi}\\
H^\phi_r &:=&\pi\phi'
\label{eq:Hphir}
\end{eqnarray}
are the energy density and momentum density of the scalar field.
The scalar field is coupled to gravity by adding the two actions:
$S^G+S^\phi$.
The total super-Hamiltonian and supermomentum are then
\begin{equation}
H:= H^G + H^\phi\,,\quad H_r:=H^G_r + H^\phi_r\,.
\end{equation}
The details leading to all of the above results can be found in
\cite{Kuchar1}.

Boundary terms and falloff conditions play an important role for vacuum
primordial black holes.
They play an equally important role for gravity coupled to a spherically
symmetric matter source.
But rather than write down the falloff conditions in all their detail, let
us just state the main results.
Namely, it is possible to choose falloff conditions on the canonical
variables $(\phi,\pi,\L,P_\L,R,P_R)$ and on the lapse and shift
$(N,N^r)$ at $r=0$ and $r\rightarrow\infty$ such that:
(i) the total action $S^G+S^\phi$ is well-defined;
(ii) the $t=const$ surfaces are free of conical singularities at
$r=0$;
(iii) no boundary terms are needed to compensate the variation of the
scalar field variables at $r=0$ and $r\rightarrow\infty$; and
(iv) no boundary terms other than
\begin{equation}
-\int dt\,N_\infty(t)M_\infty(t)
\label{eq:boundary1}
\end{equation}
are needed to compensate the variation of the gravitational variables at
$r=0$ and $r\rightarrow\infty$.
(Expression (\ref{eq:boundary1}) is equal to the boundary term at the right
infinity for the vacuum theory.
See \cite{Kuchar1} for details.)

For the boundary term (\ref{eq:boundary1}) written as above, the lapse
function cannot be freely varied at $r\rightarrow\infty$.
If it were, we would find $M_\infty(t)=0$, implying that spacetime is flat.
We can remove this restriction on the variation of the lapse by introducing
the proper time $\t_\infty$ at $r\rightarrow\infty$  as an additional
dynamical variable.
Since $N_\infty=\dot\t_\infty$, we can rewrite the boundary term as
\begin{equation}
S_{\partial\S}:=-\int dt\,\dot\t_\infty M_\infty\,.
\label{eq:boundary2}
\end{equation}
The total action is then given by
\begin{equation}
S:=S^G+S^\phi+S_{\partial\S}\,.
\end{equation}
It is to be thought of as a functional of
$(\phi,\pi,\,\L,P_\L,\,R,P_R;\,N,N^r;\,\t_\infty)$.

\section*{3. ADM mass as a canonical coordinate}

The total super-Hamiltonian and supermomentum
\begin{eqnarray}
H &=& -R^{-1} P_R P_\L+ {1\over 2} R^{-2} \L P_\L{}^2\nonumber\\
& & \mbox{} + \L^{-1}RR'' - \L^{-2}RR'\L' +
{1\over 2} \L^{-1} R'^{2} - {1\over 2}\L \nonumber\\
& & \mbox{} +{1\over 2}\L^{-1}\left(R^{-2}\pi^2+R^2\phi^2\right)
\label{eq:Htot}\\
H_r&=&P_R\,R'-\L P_\L{}'+\pi\phi'\label{eq:Htotr}
\end{eqnarray}
are complicated expressions of the gravitational variables
$(\L,P_\L,R,P_R)$.
We desire a canonical transformation to a new set of variables, in terms of
which the constraints $H=0=H_r$ simplify.\footnote
{Alternatively, one may choose to solve the constraints $H=0=H_r$ by
imposing the coordinate and slicing conditions $r=R,\,P_\L=0$.
This is the approach followed by BCMN\cite{BCMN}, Unruh\cite{Unruh},
and H\'aj\'{\i}\v{c}ek\cite{Hajicek} in their papers.
We will not follow their approach here, since we do not want to restrict
ourselves to a privileged foliation.}

For vacuum spherically symmetric spacetimes, Kucha\v{r}\cite{Kuchar1} found
such a canonical transformation.
He showed that the mapping $(\L,P_\L,R,P_R)\mapsto(M,P_M,\sR,P_\sR)$
given by
\begin{eqnarray}
M &:=& {1\over 2}R^{-1}P_\L^2-{1\over 2}\L^{-2}R R'{}^2+{1\over 2}R
\label{eq:M}\\
P_M &:=& R^{-1}F^{-1}\L P_\L\label{eq:PM}\\
\sR &:=& R\label{eq:sR}\\
P_\sR &:=& P_R - {1\over 2}R^{-1}\L P_\L-{1\over 2}R^{-1}F^{-1}\L P_\L
\nonumber\\
& &-R^{-1}\L^{-2}F^{-1}\left(\,(\L P_\L)'(RR')-(\L P_\L)(RR')'\,\right)
\label{eq:PsR}
\end{eqnarray}
where
\begin{equation}
F := \left({R'\over\Lambda}\right)^2 - \left({P_\L\over R}\right)^2
\end{equation}
is a canonical transformation on the gravitational phase space
{\it irrespective of constraints or dynamics}.
As such, it remains a canonical transformation on the extended phase space
that includes the scalar field variables $(\phi,\pi)$.

In terms of the new canonical variables, the expressions for the
super-Hamiltonian and supermomentum simplify considerably:
\begin{eqnarray}
\Lambda H &=& -F^{-1}M'\sR' - F P_M P_\sR
+{1\over 2}\left(\sR^{-2}\pi^2 + \sR^2\phi'{}^2\right)\label{eq:Hnew1}\\
H_r &=&P_\sR\sR' + P_M M' + \pi\phi'\label{eq:Hrnew1}
\end{eqnarray}
where
\begin{equation}
F=1-2M/\sR\,.
\end{equation}
Notice that the left hand side of (\ref{eq:Hnew1}) is the product of
$\L\not=0$ and $H$.
We are allowed to perform such a scaling without changing the constraint
$H=0$.

For vacuum spherically symmetric spacetimes, the canonical coordinate
$M(r)$ is the Schwarzschild mass of the spacetime.
In fact, Kucha\v{r}\cite{Kuchar1} obtained expression (\ref{eq:M}) for
$M(r)$ by equating the ADM form of the spacetime line element (constructed
from $N$, $N^r$, and $d\sigma$) with the Schwarzschild line element
\begin{equation}
ds^2=-\left(1-{2M\over R}\right)\,dT^2+\left(1-{2M\over R}\right)^{-1}
\,dR^2+R^2\,d\Omega^2
\label{eq:ds2schw}
\end{equation}
for an arbitrary parametrization: $T=T(t,r),\,R=R(t,r)$.
It turns out that this reconstruction program for the mass also works for
gravity coupled to an arbitrary, spherically symmetric, matter source.
Instead of (\ref{eq:ds2schw}), we have
\begin{equation}
ds^2=-G(T,R)\,dT^2+\left(1-{2M(T,R)\over R}\right)^{-1}
\,dR^2+R^2\,d\Omega^2
\label{eq:ds2}
\end{equation}
where $G(T,R)$ is in general different from
\begin{equation}
F(T,R):=1-2M(T,R)/R\,.
\end{equation}
As shown, for example, by Synge\cite{Synge} and Thorne\cite{Thorne},
$M(T,R)$ equals the total ADM mass of the spacetime contained within the
sphere of curvature radius $R$ at the time $T$.
Thus, the canonical coordinate $M(r)$ has a good physical meaning for any
spherically symmetric matter source coupled to gravity.
(See also the papers by Guven and \'O Murchadha\cite{GuvenOM}.)

\section*{4. Spacetime problem of time}

For vacuum spherically symmetric spacetimes, the introduction of the
Schw\-arzschild mass as a canonical variable served only as an intermediate
step.
After carefully taking into account the boundary terms at the left and
right infinities, Kucha\v{r}\cite{Kuchar1} subsequently performed a
transformation that turned the Killing time $T$ of the Schwarzschild
solution into a canonical coordinate $\sT(r)$ on the geometrodynamical
phase space, and then solved the constraints.
As mentioned in Sec.~1, the final result is extremely simple:
\begin{equation}
P_\sT(r)=0\,,\quad P_\sR(r)=0\,.
\end{equation}

For gravity coupled to a spherically symmetric matter source, the same
transformation (modified slightly to account for the different topology of
$\S$) can be performed.
Unfortunately, the final result for this case is not nearly as nice.
First, the constraints $H=0=H_r$ do not lend themselves to any obvious
solution.
Second, even if we could solve the constraints for the momenta canonically
conjugate to $\sT(r)$ and $\sR(r)$, $\sT(r)$ is no longer a spacetime scalar
once spherically symmetric matter is coupled to gravity.
Thus, this solution of the super-Hamiltonian and supermomentum
constraints---even if it exists---suffers from the so-called
{\it spacetime problem of time}\cite{Kuchar2}.

Let us be more specific.
Consider the transformation $(\t_\infty,M,P_M)\mapsto(\sT,P_\sT)$ given by
\begin{eqnarray}
\sT(r) &:=& \t_\infty - \int_\infty^r d\bar{r}\,P_M(\bar{r})\,,
\label{eq:sT}\\
P_\sT(r) &:=& -M'(r)\,.
\label{eq:PsT}
\end{eqnarray}
This is Kuchar's canonical transformation adapted to the topology
$\S={\rm I\!R}^3$.
The mapping (\ref{eq:sT})--(\ref{eq:PsT}) is invertible:
\begin{eqnarray}
M(r) &=& - \int_0^r d\bar{r}\,P_\sT(\bar{r})\,,\label{eq:Minv}\\
P_M(r) &=& -\sT'(r)\,,\label{eq:PMinv}\\
\t_\infty &=& \sT(\infty)\,.\label{eq:tauinftyinv}
\end{eqnarray}
It also sends
\begin{equation}
\int_0^\infty dr\,P_M(r)\,\delta M(r)-M_\infty\,\delta\t_\infty\,
\mapsto\,
\int_0^\infty dr\,P_\sT(r)\,\delta\sT(r)
\end{equation}
modulo an exact differential.
Thus, $(\phi,\pi,\sR,P_\sR,\sT,P_\sT)$ is a canonical chart on the extended
phase space.

In terms of these new variables, the super-Hamiltonian and supermomentum
are given by
\begin{eqnarray}
\Lambda H &=& F^{-1}P_\sT\sR' + F\,P_\sR\sT'
+ {1\over 2}\left(\sR^{-2}\pi^2+ \sR^2\phi'{}^2\right)
\label{eq:Hnew2}\\
H_{r} &=& P_\sR\sR' + P_\sT\sT' + \pi\phi'
\label{eq:Hrnew2}
\end{eqnarray}
where
\begin{equation}
F(r)=1+{2\over\sR(r)}\int_0^r\,d\bar{r}\,P_\sT(\bar{r})\,.
\end{equation}
Although these expressions for $\L H$ and $H_r$ are much simpler than they
were originally (see Eqs.~(\ref{eq:Htot}) and (\ref{eq:Htotr})), it is
still not obvious how to solve the constraints $H=0=H_r$.
The culprits are the $F^{-1}$ and $F$ factors multiplying the first two
terms of the scaled super-Hamiltonian (\ref{eq:Hnew2}).
These factors are responsible for the nonlinear dependence of $\L H$ on
$P_\sT$.
We did not succeed to solve these equations for $P_\sT$ and $P_\sR$ on a
general hypersurface.\footnote
{If we choose to impose the coordinate and slicing conditions $r=\sR,\,
\sT'=0$ (which are equivalent to the BCMN gauge conditions
$r=R,\,P_\L=0$---see footnote 1), we {\it can} solve the constraints for
the momenta $P_\sT$ and $P_\sR$, and recover the BCMN-Unruh reduced
Hamiltonian.
The solution is
\begin{equation}
P_\sT(\sR) = -{d\over d\sR}\,\left(\,{\sR\over 2}\left(1-|F|\right)
\,\right)\,,\quad
P_\sR(\sR) = -\pi(\sR)\phi'(\sR)\nonumber
\end{equation}
where
\begin{equation}
|F(\sR)|=\sR^{-1}\,\exp\bigg(-\int_\infty^\sR\,{\cal S}_\phi\bigg)
\,\int_0^\sR d\bar{\sR}\, \exp\bigg(\int_\infty^{\bar{\sR}}\,
{\cal S}_\phi\bigg)
\nonumber
\end{equation}
and
\begin{equation}
{\cal S}_\phi(\sR):=\sR^{-1}\left(\sR^{-2}\pi^2+\sR^2\phi'{}^2\right)\,.
\nonumber
\end{equation}
The Hamiltonian for the reduced theory is simply
\begin{equation}
H_{\rm red}[\phi,\pi]=-\int_0^\infty d\sR\,P_\sT(\sR)
=-{1\over 2}\int_0^\infty d\sR\,\left[\,\exp\left(\int_\infty^\sR
\,{\cal S}_\phi\right)-1\right]\nonumber
\end{equation}
which agrees (up to a factor of ${1\over 4}$) with the BCMN-Unruh reduced
Hamiltonian.
This calculation just serves as a check on our results.
As mentioned in Sec.~1, we prefer not to work on a privileged foliation.}
\vfill\eject

But let us suppose, for the sake of argument, that we were able somehow
to solve the constraints for $P_\sT$ and $P_\sR$.
Could we then claim that we found a satisfactory functional time formalism
for a spherically symmetric, massless, scalar field coupled to gravity?
The answer is ``no.''
The reason is the following:
A true embedding variable must be a spacetime scalar; it should not depend
on the hypersurface from which it was constructed.
If two hypersurfaces $\S$ and $\S'$ intersect at the same event ${\cal E}$
in spacetime, and if the canonical data on each of these hypersurfaces are
related by the Einstein equations, then the values of the embedding variable
at ${\cal E}$ (obtained from the two sets of canonical data) must be equal.
Otherwise, the embedding variable would assign different values to the same
spacetime point.
Since, as we shall show below, $\sT(r)$ is {\it not} a spacetime scalar,
it is not a true embedding variable.
We do not have a functional time formalism for our theory, and this solution
suffers from the spacetime problem of time\cite{Kuchar2}.

The requirement that a dynamical variable be a spacetime scalar can be
expressed in purely canonical language\cite{Kuchar2}.
Namely, a dynamical variable $s(x)$ is a spacetime scalar if and only if:
(i) the function $s(x)$ is a spatial scalar; and
(ii) the value $s(x)$ is unchanged, modulo the constraints, if we
evolve the canonical data with a smeared super-Hamiltonian whose smearing
function vanishes at $x$.
Condition (ii) is equivalent to
\begin{equation}
\{s(x),H(y)\}\approx\,\propto\delta(x,y)
\label{eq:EH}
\end{equation}
where $\propto\delta(x,y)$ is shorthand notation for terms proportional
to $\delta(x,y)$.
Thus, the Poisson bracket of a spacetime scalar with the super-Hamiltonian
is weakly proportional to a $\delta$-function.

A fairly simple calculation shows that condition (\ref{eq:EH}) is not
satisfied for $\sT(r)$.
Explicitly,
\begin{equation}
\{\sT(r),\L(\bar{r})H(\bar{r})\}\approx F^{-1}(\bar{r})\,
\sR'(\bar{r})\,\delta(r,\bar{r}) + \propto\Theta(\bar{r}-r)
\end{equation}
where the coefficients of the terms multiplying the step function
$\Theta(\bar{r}-r)$ are not weakly equal to zero unless the scalar field
vanishes.
Thus, $\sT(r)$ is not a spacetime scalar, and this solution of the
constraints---even if it exists---suffers from the spacetime problem of
time.

To conclude this section, we point out that, modulo certain technical
difficulties\footnote
{These amount to the non-invertibility of the transformation
(\ref{eq:bT})--(\ref{eq:PbR}).
We can only recover the absolute value of $F$ if we try to invert the
the transformation.
(See Eq.~(\ref{eq:absF}).)
If we try to avoid this problem by restricting ourselves to one sign of $F$,
say $F(r)>0$, then we lose the hypersurfaces which penetrate an apparent
horizon $F(r)=0$.},
we can perform a transformation to a new time variable $\bT(r)$ in terms of
which we can explicitly solve the constraints.
The transformation $(\t_\infty,M,P_M,\sR,P_\sR)\mapsto(\bT,P_\bT,\bR,P_\bR)$
is
\begin{eqnarray}
\bT(r) &:=& \t_\infty - \int_\infty^r d\bar{r}\,F(\bar{r})P_M(\bar{r})
\label{eq:bT}\\
P_\bT  &:=& {d\over dr}\,\left(\,{\sR\over 2}\,\ln|F|\,\right)
\label{eq:PbT}\\
\bR    &:=& \sR\label{eq:bR}\\
P_\bR  &:=& P_\sR+{1\over 2}F\,P_M\,\left(F^{-1}-1+\ln|F|\right)
\label{eq:PbR}
\end{eqnarray}
where
\begin{equation}
F=1-2M/\sR\,.
\end{equation}
The transformed constraints are
\begin{eqnarray}
\Lambda H &=& P_\bT\bR' + P_\bR\bT' + {1\over 2}\left(\bR^{-2}\pi^{2}+
\bR^2\phi'{}^{2}\right)\nonumber\\
& & -{1\over 2}\left(\bR'{}^2-\bT'{}^2\right)\left(F^{-1}-1+\ln|F|\right)
\label{eq:Hnew3}\\
H_r &=& P_{\bR}\bR'+P_{\bT}\bT' + \pi\phi'
\label{eq:Hrnew3}
\end{eqnarray}
where
\begin{equation}
|F(r)|=\exp\left(\,{2\over \bR(r)}\int_0^r\,d\bar{r}\,P_\bT(\bar{r})\,
\right)\,.
\label{eq:absF}
\end{equation}
The solution of the constraints is
\begin{eqnarray}
P_\bT(r)&=&{d\over dr}\,\left({\bR(r)\over 2}\ln|F(r)|\right)\\
P_\bR(r)&=&-{1\over \bR'(r)}\,\bigg(P_\bT(r)\bT'(r)+\pi(r)\phi'(r)
\bigg)
\end{eqnarray}
where
\begin{equation}
|F(r)|=\bR^{-1}(r)\,\exp\bigg(-\int_\infty^r \,{\cal S}_\phi\bigg)
\,\int_0^r d\bar{r}\,\bR'(\bar{r})\,\exp\bigg(\int_\infty^{\bar{r}}
\,{\cal S}_\phi\bigg)
\end{equation}
and
\begin{equation}
{\cal S}_\phi(r):=-2\,\bR^{-1}(\bR'{}^2-\bT'{}^2)^{-1}\left(\bT'\, \pi\phi'
-{1\over 2}\bR'\left(\bR^{-2}\pi^2+\bR^2\phi'{}^2\right)\right)\,.
\end{equation}
If we impose the coordinate and slicing conditions $r=\bR,\,\bT'=0$, our
solution again reproduces the BCMN-Unruh reduced Hamiltonian.
(See footnote 2.)

Unfortunately, just like $\sT(r)$, $\bT(r)$ is not a spacetime scalar:
\begin{equation}
\{\bT(r),\L(\bar{r})H(\bar{r})\}\approx\bR'(\bar{r})\,
\delta(r,\bar{r})+\propto\Theta(\bar{r}-r)
\end{equation}
where again the coefficients of the terms multiplying the step function
are not weakly equal to zero.
Thus, this explicit solution of the constraints also suffers from the
spacetime problem of time.

\section*{5. Discussion}

The time variables $\sT(r)$ and $\bT(r)$ that we introduced as canonical
coordinates on phase space both failed to be spacetime scalars.
As such, they did not qualify as true embedding variables.
It is important to stress, however, that the two attempts presented in this
paper {\it do not constitute a proof} that a functional time formalism for
spherically symmetric matter systems coupled to gravity does not exist.
In fact, as we shall argue below, our current belief is that a functional
time formalism for these systems {\it does} exist.
We need only be more clever in our choice of time variable.\footnote
{A functional time formalism for spherically symmetric spacetimes has also
been discussed by Braham\cite{Braham}.
It appears that his solution of the constraints also suffers from the
spacetime problem of time.
The ``embedding'' variables given in \cite{Braham} are not spacetime scalars.}

Indeed, a natural candidate for a time variable that {\it is} a spacetime
scalar is the curvature time of the general, spherically symmetric,
spacetime line element
\begin{equation}
ds^2 = - G(T,R)\,dT^2 + F(T,R)^{-1}\,dR^2 + R^2\,d \Omega^{2}
\label{eq:ds2a}
\end{equation}
where
\begin{equation}
F(T,R):=1-2 M(T,R)/R\,.
\label{eq:ds2b}
\end{equation}
(See also the discussion at the end of Sec.~3.)
By its definition, the curvature time $T$ is a spacetime scalar, and like
the curvature radius $R$, $T$ has an invariant geometrical meaning:
(i) the surfaces of constant $T$ are orthogonal to the lines of constant
$R$, $\theta$, and $\phi$; and
(ii) the labeling of the $T=const$ surfaces is specified (up to the choice
of time origin) by requiring that $T$ measure proper time at $R=0$.
Requirement (ii) imposes the boundary condition $G(T,R=0)=1$ on $G$.

The problem is how to turn this privileged spacetime coordinate into a
canonical coordinate on our phase space.

For vacuum spherically symmetric spacetimes, there is no problem.
Following the reconstruction program for the mass described in Sec.~3, one
finds
\begin{equation}
-T'=F^{-1}R^{-1}\L P_\L
\label{eq:vacT'}
\end{equation}
where
\begin{equation}
F=\left({R'\over\L}\right)^2 - \left({P_\L\over R}\right)^2\,.
\label{eq:F5}
\end{equation}
As shown in \cite{Kuchar1}, $-T'(r)$ is the momentum canonically conjugate
to the Schw\-arzschild mass $M(r)$.
Then, by carefully taking into account the boundary terms at the left
and right infinities, one can perform another transformation that turns
$T$ itself into a canonical coordinate on phase space.
(See \cite{Kuchar1} for more details.)

For gravity coupled to a spherically symmetric matter source, things are not
so simple.
Equation (\ref{eq:vacT'}) is replaced by
\begin{equation}
-G^{1\over 2}T'=F^{-{1\over 2}}R^{-1}\L P_\L
\end{equation}
where $F$ is given by our old expression (\ref{eq:F5}).
Thus, we have only been able to reconstruct the product $G^{1\over 2}T'$ in
terms of the original gravitational variables.
To obtain an expression for $G$ or $T'$ separately, we must somehow involve
the matter variables.

An idea that immediately suggests itself is to use one of the Einstein
equations\cite{Synge}:
\begin{equation}
G(T,R)=F(T,R)\,\exp\bigg[\,8\pi\int_0^R d\bar{R}\, \bar{R}\,F^{-1}
(T,\bar{R})\,\Big(\mu(T,\bar{R})+p(T,\bar{R})\Big)\,\bigg]
\label{eq:GfromF}
\end{equation}
where
\begin{equation}
\mu:=-{\cal T}_T^T\,,\quad p:={\cal T}_R^R
\end{equation}
are two components of the energy-momentum tensor ${\cal T}_{\alpha\beta}$
for the spherically symmetric matter source.
The problem with this approach is that the integral in (\ref{eq:GfromF})
is over a $T=const$ surface.
Even though it is possible to express the integrand of (\ref{eq:GfromF}) in
terms of the original canonical variables, we still have to evolve the
canonical data from $\S$ to the $T=const$ surface before we can do the
integration.
Since $\S$ is an arbitrary spherically symmetric hypersurface, $\S$ need
not agree with the $T=const$ surface anywhere.
The resulting expression for $G$, and hence for $T$, would be non-local in
time as well as in space.

Another approach, which appears to be more promising, has a somewhat
different starting point.
The idea is to first reduce the Einstein-Hilbert action to spherically
symmetric spacetime metrics of the form (\ref{eq:ds2a}) and (\ref{eq:ds2b}),
and then parametrize the resulting action to introduce the curvature time
$T$ and its conjugate momentum as canonical data on arbitrary, spherically
symmetric, hypersurfaces.
In this manner, we would succeed in promoting both the curvature radius $R$
and curvature time $T$ to canonical coordinates $R(r)$ and $T(r)$ on phase
space.
The spacetime problem of time would thereby be avoided.
But a possible problem with this approach is the existence of {\it second
class constraints}.
In the process of eliminating the second class constraints prior to
quantization, we may lose $T(r)$ as one of our canonical variables.
We are currently investigating these issues.

\section*{Acknowledgments}

I would like to thank Karel Kucha\v{r} for suggesting this problem, and
for his many insightful comments and questions.
I would also like to thank Carsten Gundlach and Don Marolf for discussions
during the initial stages of this work.
This research was supported in part by the NSF grants PHY89-04035 and
PHY-9207225, and by the U.S.-Czech Science and Technology Grant No~92067.

\end{document}